\documentclass[]{pasj01}

\usepackage{natbib,aas_macros}

\begin{document} 
\Received{2019/6/19}
\Accepted{2019/10/16}

\title{Near-infrared Monitoring of the Accretion Outburst in the MYSO S255-NIRS3}

\author{Mizuho \textsc{uchiyama}\altaffilmark{1}}%
\email{mizuho.uchiyama@nao.ac.jp}

\author{Takuya \textsc{yamashita}\altaffilmark{2}}
\author{Koichiro \textsc{sugiyama}\altaffilmark{3,4}}

\author{Tatsuya \textsc{nakaoka}\altaffilmark{5}}
\author{Miho \textsc{kawabata}\altaffilmark{6,7,5}}
\author{Ryosuke \textsc{itoh}\altaffilmark{8,5}}
\author{Masayuki \textsc{yamanaka}\altaffilmark{6,7,5}}
\author{Hiroshi \textsc{akitaya}\altaffilmark{5}}
\author{Koji \textsc{kawabata}\altaffilmark{5}}

\author{Yoshinori \textsc{yonekura}\altaffilmark{9}}
\author{Yu \textsc{saito}\altaffilmark{9}}
\author{Kazuhito \textsc{motogi}\altaffilmark{10}}
\author{Kenta \textsc{fujisawa}\altaffilmark{11}}

\altaffiltext{1}{Advanced Technology Center, National Astronomical Observatory of Japan (NAOJ), 2-21-1 Osawa, Mitaka, Tokyo 181-8588, Japan}
\altaffiltext{2}{TMT-J Project Office, NAOJ, 2-21-1 Osawa, Mitaka, Tokyo 181-8588, Japan}
\altaffiltext{3}{Mizusawa VLBI Observatory, NAOJ, 2-21-1 Osawa, Mitaka, Tokyo 181-8588, Japan}
\altaffiltext{4}{National Astronomical Research Institute of Thailand, 260 Moo 4, T.Donkaew, A. Maerim, Chiang Mai, 50180, Thailand}
\altaffiltext{5}{Hiroshima Astrophysical Science Center, Hiroshima University, 1-3-1 Kagamiyama, Higashi-Hiroshima, Hiroshima 739-8526, Japan}
\altaffiltext{6}{Department of Astronomy, Kyoto University, Kitashirakawa-Oiwake-cho, Sakyo-ku, Kyoto 606-8502, Japan}
\altaffiltext{7}{Okayama Observatory, Kyoto University, 3037-5 Honjo, Kamogata, Asakuchi, Okayama 719-0232, Japan}
\altaffiltext{8}{Bisei Astronomical Observatory, 1723-70 Okura, Bisei, Ibara, Okayama 714-1411, Japan}
\altaffiltext{9}{Center for Astronomy, Ibaraki University, 2-1-1 Bunkyo, Mito, Ibaraki 310-8512, Japan}
\altaffiltext{10}{Graduate School of Sciences and Technology for Innovation, Yamaguchi University, 1677-1 Yoshida, Yamaguchi, Yamaguchi 753-8512, Japan}
\altaffiltext{11}{The Research Institute for Time Studies, Yamaguchi University, 1677-1 Yoshida, Yamaguchi, Yamaguchi 753-8511, Japan}


\maketitle

\begin{abstract}
	
We followed-up the massive young stellar object (MYSO) S255-NIRS3 (=S255-IRS1b) during its recent accretion outburst event
in the \textsl{Ks} band with
Kanata/HONIR for four years after its burst and obtained a long-term light curve.
This is the most complete NIR light-curve of the S255-NIRS3 burst event that has ever been presented.
The light curve showed a steep increase reaching a peak flux
that was 3.4 mag brighter than the quiescent phase and then a relatively moderate year-scale fading until the last observation,
similar to that of the accretion burst events such as EXors found in lower-mass young stellar objects.
The behavior of the \textsl{Ks} band light curve is similar to that observed in 6.7 GHz class II methanol maser emission,
with a sudden increase followed by moderate year-scale fading. However, the maser
emission peaks appear 30-50 days earlier than that of the \textsl{Ks} band emission. 
The similarities confirmed that the origins of the maser emission and the \textsl{Ks} band continuum emission is common as previously shown from another infrared and radio observations by \citet{Stecklum16, Caratti17, Moscadelli17}.
However, the differences in energy transfer paths, such as the exciting/emitting/scattering
structures, may cause the delay in the flux-peak dates.

\end{abstract}

\section{Introduction}
Massive stars play an important role in the galaxies and cosmic evolution.
Their feedbacks, such as supernova explosions and stellar winds, affect the surrounding environment significantly \citep{Tan14, ZY07}.
However, detailed mass-growth processes of massive young stellar objects (MYSOs) were not clear until now.
Three major candidate processes---core accretion, competitive accretion, and stellar collision---have been proposed but it is still not clear which process is the dominant one although stellar collision might be efficient only in an extremely crowded region and might not be a frequent process \citep{Tan14}.
Because the formation of massive stars is deeply obscured by dense gas and dust, and is located relatively far from us (typically farther than 1 kpc), their physical structure is hard to resolve.
In fact, currently available spatial resolution of observations is insufficient to resolve inner 10AU-scale structures of the MYSOs and their disks, where the mass growth phenomenon happens, unless VLBI technique are used \citep{Tan14}.
VLBI observations, however, mostly detect only non-thermal emissions such as masers, and it is difficult to investigate the physical parameters and structures of the emitting regions. Therefore, other observational methods are required to observe the inner structures in more detail.

Variability study is a promising method for this purpose.
Such studies of low-mass young stellar objects at optical and infrared wavelengths have proven to be powerful tools for deciphering the physics of star formation and pre-main-sequence stellar evolution, such as stellar rotation, the existence and characteristics of cool/hot spots, infalling/rotating dusty objects, and the variation in accretion rate \citep{Morales11}.

Although the peak of spectral energy distribution in MYSOs is located in the mid-infrared (MIR) or the far-infrared (FIR), near-infrared (NIR) observations are suitable for long-term and/or high-cadence monitoring owing to accessibility of the available facilities.
One pioneering study reported the observation of seven reflection nebulae of Cep A East in the \textsl{Ks} band for several years and showed the variability of all the components \citep{Hodapp09}. The authors suggested that the variability should be caused by the variation in dust extinction caused by a rotating structure around $\sim 10$ AU from the MYSO \citep{Hodapp09}.
Recently, 13 high amplitude ({$\Delta \textsl{K} > 1$~mag}) year-scale variable MYSOs in the \textsl{Ks} band were reported for the first time using VVV survey data \citep{Kumar16} and another work also detected 190 lower-amplitude variable MYSOs out of 718 samples \citep{Teixeira18}.
These studies imply that variability is common in MYSOs also and have the potential to reveal the physical processes happening in the inner regions of MYSOs.
Higher-cadence observations of these variable MYSOs are required to get more details of their physical structure and the phenomena happening much inside the MYSOs and/or their surrounding disks.
In addition to such observations, NIR follow-up observations of possible drastic changes in the physical conditions of such MYSOs, such as sudden increase in radio fluxes, are also important to study their nature.

Among the origins of variability, accretion burst events are well-known events causing drastic variability in low-mass young stellar objects (YSOs) \citep[][and renfenreces there in]{Audard14}.
Recently, their appearance was theoretically proposed in MYSOs as well \citep{Meyer17} and observed for the first time in S255-NIRS3 at infrared wavelengths \citep{Caratti17}.
Later on, another accretion burst in an MYSO (NGC 6334I-MM1) was detected at sub-mm wavelengths \citep{Hunter17}.
In addition, the radio jet burst triggered by the accretion burst was also detected in S255-NIRS3 \citep{Cesaroni18}.

In October 2015, a class II methanol maser outburst at 6.7 GHz was reported in S255 region \citep{Fujisawa15}.
This discovery triggered a search for the exciting source.
Because a class II methanol maser is expected to be associated with the thermal emission of dust in disks \citep{Norris93,Minier00,Cragg05,Bartkiewicz09,Sugiyama14}, origins of the maser outburst would be very close to the MYSO, and flux variation in the infrared would also be expected.
\citet{Stecklum16} identified S255 NIRS3, with a trigonometric distance of 1.78 kpc \citep{Burns16}, as the exciting source suggesting, for the first time, the possibility of an accretion outburst in a massive YSO and its connection with the maser flare. 
Based on their infrared imaging and spectroscopic study, \citet{Caratti17} indicated that this event was triggered by an accretion burst due to a five-fold increase in the luminosity and intensity of emission lines related to mass accretion 
and this was confirmed by following VLBI observations by \citet{Moscadelli17}.

Note that in some previous NIR imaging studies of the S255 region \citep{Ojha11}, the authors named this object as S255-IRS1b, referred from \citet{Howard97}, while some NIR and radio observational works released after the burst detection \citep{Caratti17,Moscadelli17,Cesaroni18} named it as S255-NIRS3, referred from \citet{Tamura91}.
In this work, we refer to this MYSO as S255-NIRS3.

The previous NIR observations by \citet{Stecklum16} and \citet{Caratti17} clearly showed the drastic flux increase in S255-NIRS3.
However, they only reported two monitoring observations and the NIR peak flux of the S255-NIRS3 during the maser burst was missing. Furthermore, their monitoring duration was approximately six months and a whole picture of long-term variability behavior of S255-NIRS3 was not clear in the NIR.
In this study, we present the first \textsl{Ks} band photometric follow-up observations of the accretion burst and analyze the NIR light-curve trend with respect to the methanol maser light-curves at several major velocities.
We conducted high-cadence imaging monitoring of S255-NIRS3 in the \textsl{Ks} band and made a detailed light curve for as long as four years, clearly showing week- to year-scale variability and the flux peak after the maser outburst event.
We also compare our result with high-cadence observations of the 6.7 GHz methanol maser light-curve from other studies \citep[][Sugiyama et al. in prep.]{Szymczak18} and discuss their relationship.

\section{Observations}

\subsection{NIR monitoring}

\begin{figure}[p] 
  \begin{center}
    \includegraphics[width=8cm]{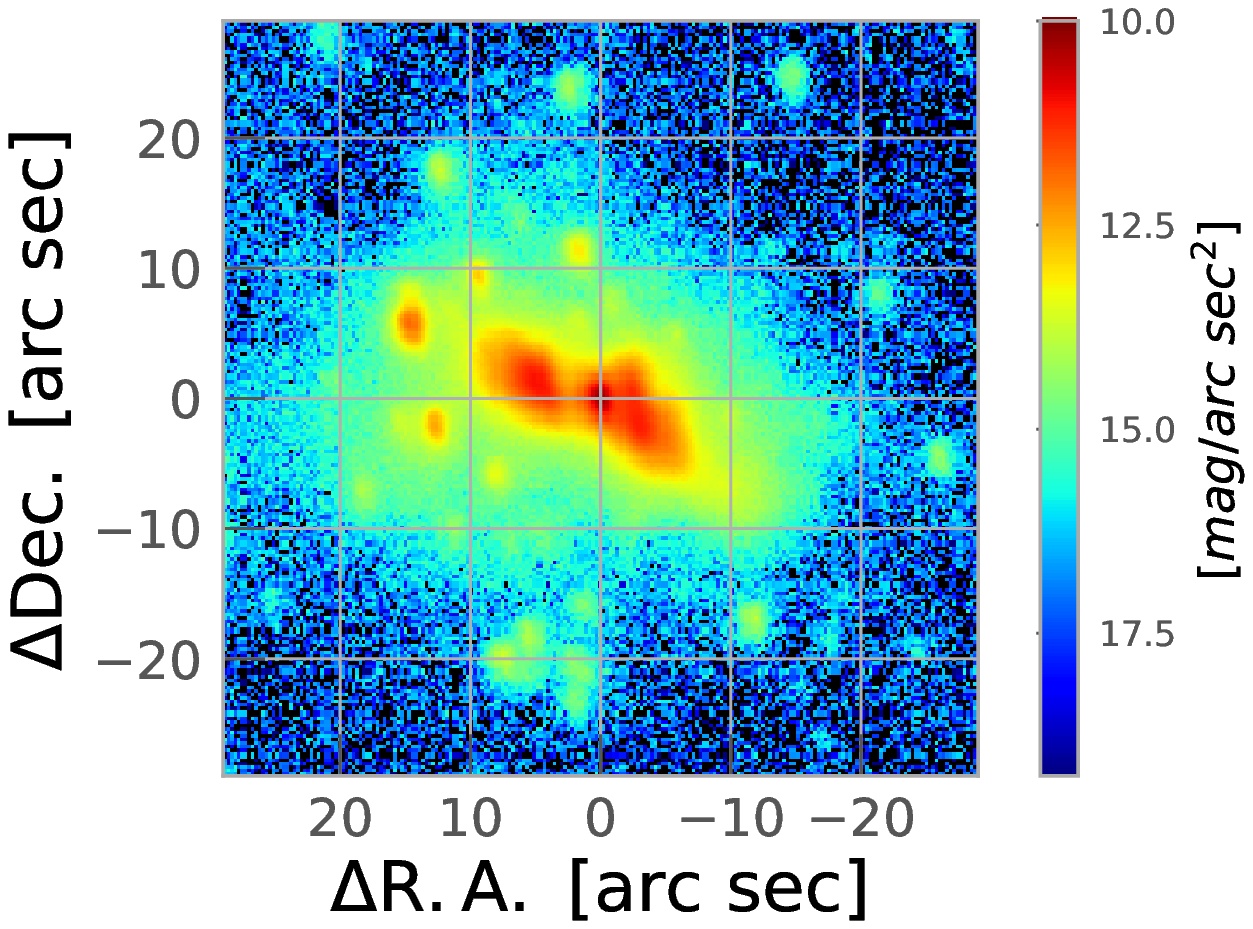}\\
    \includegraphics[width=8cm]{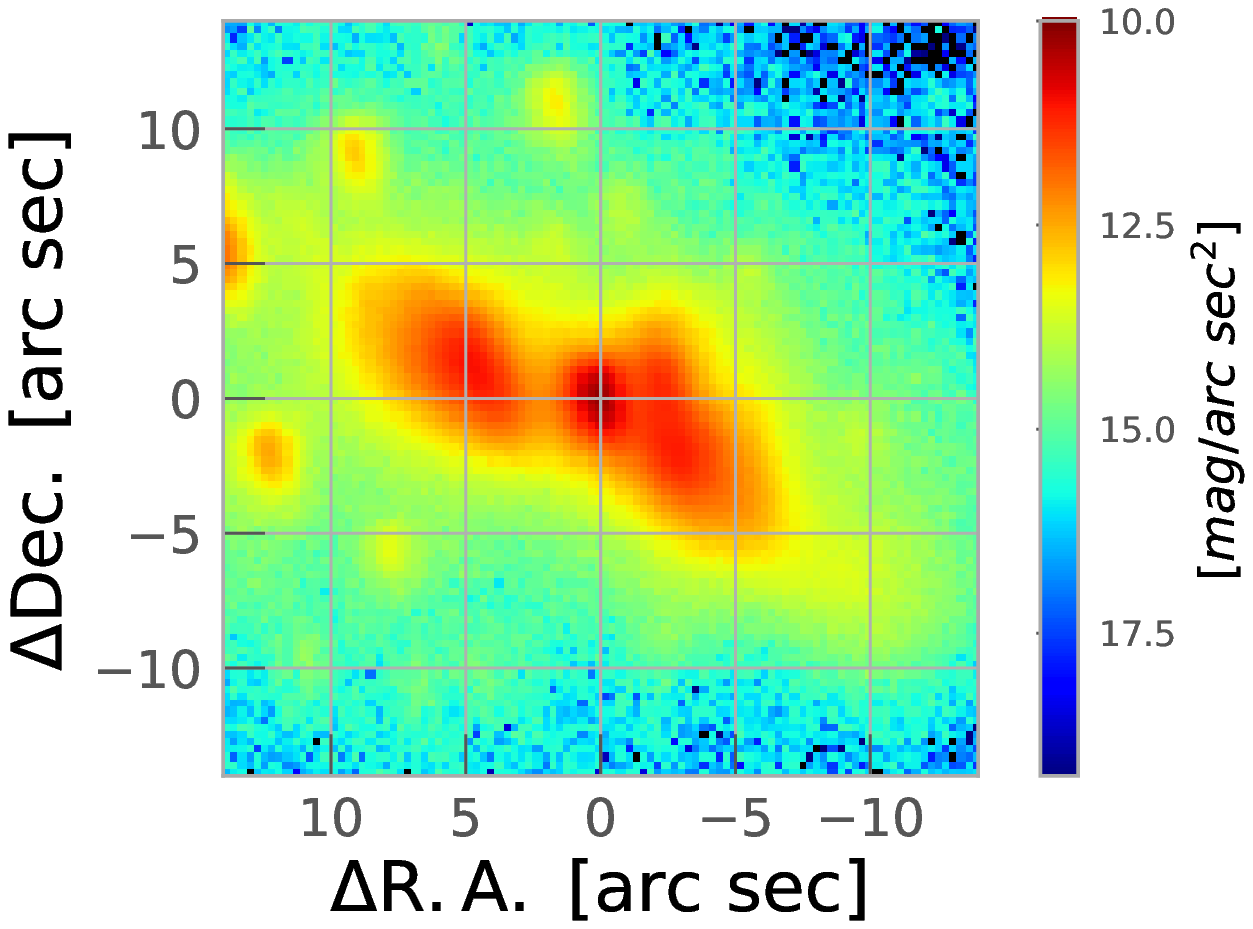}\\
    \includegraphics[width=6cm]{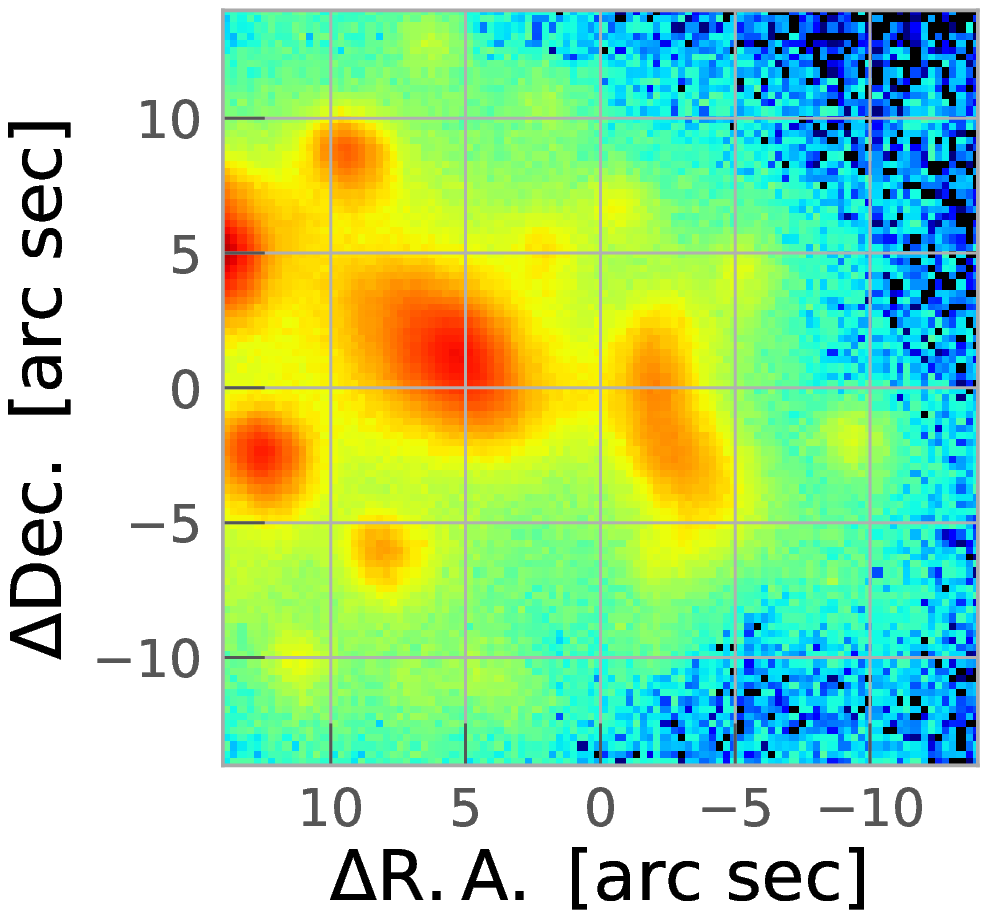}\\
  \end{center}  
\caption{(Top) Image of S255-NIRS3 region observed with Kanata/HONIR in the \textsl{Ks} band on December 1, 2015. (Middle) The same image but zoomed up. (Bottom) The same as middle but in the \textsl{H} band. The center positions are \timeform{06h12m54d.020}, \timeform{+17D59'23''.07} in J2000.0, corresponding to S255-NIRS3 within the seeing. The position of the known methanol maser spot is located about 0.1 arcsec east to S255-NIRS3. }\label{fig:field}
\end{figure}

Monitoring observations in the NIR were conducted with HONIR installed on the 1.5 m Kanata telescope located at the Higashi-Hiroshima Observatory in Japan. Although HONIR can perform simultaneous optical and NIR observations, only NIR imaging data were used in this work owing to invisibility of the object in the optical.
HONIR has an infrared detector of Virgo-2K with a pixel scale of 0.295 arcsec/pix and a field of view of approximately 10 arcminutes square.
A detailed performance of HONIR was reported in \citet{Akitaya14}.
Monitoring observations were conducted in the \textsl{Ks} band on 31 nights from November 29, 2015 to March 27, 2019 with blank periods in the spring and summer of 2016, 2017, and 2018.
Exposure time of each observation is 20 seconds and five-point-dithering are performed on each observing date.
A world coordinate system was defined using registration of field stars and the coordinate of the burst object in the \textsl{Ks} band (\timeform{06h12m54d.09}, \timeform{+17D59'23''.1}) corresponded to that of S255-NIRS3 (\timeform{06h12m54d.020}, \timeform{+17D59'23''.07}; \citet{Stecklum16}) within average seeing size of 2.5 arcsec during the whole observations. 
Observations in the \textsl{H} band were also performed.
A diffuse and faint object is detected at coordinates of S255-NIRS3 (See Figure \ref{fig:field} Bottom). 
However, relatively bad seeing as compared to the previous NIR observations \citep{Stecklum16,Caratti17} and contamination from surrounding nebulae made it hard to perform PSF fitting photometry and light-curve in the \textsl{H} band is not available in this work.

\subsection{Data reduction and photometry}
Basic data reduction was conducted with a standard reduction pipeline of HONIR based on the IRAF package.
We performed point spread function (PSF) fitting photometry using the DAOPHOT package in the IRAF package.
Note that we did not use data with relatively bad seeing, Full Width Half Maximum of PSF larger than 3 arcsec, to improve photometric accuracy.

We also picked the most stable pair of stars in the field with the following method.
First, we measured the ADU counts of all the stars in the field whose counts were larger than a tenth of that of S255-NIRS3 with the same PSF fitting photometry method. Second, pairs of stars were made among all the counted stars and a count ratio of each pair was derived during the whole observation period; and finally, the most stable pair among them, whose count ratio showed a RMS variation of just 5.9 percent, equal to 0.062 magnitudes (See Figure \ref{fig:lc} and Table \ref{tab:phot}), throughout the whole monitoring program, were picked as the reference stars. 
The fluxes of S255-NIRS3 were derived and calibrated by comparing them with the brighter star in the stable pair, 2MASS J06125939+1757214 (J=10.6 mag, H=10.1 mag, K=9.9 mag), whose magnitudes are taken from the 2MASS catalog \citep{2MASS06}.
We also derived photometric error during each night from standard deviation of magnitudes of the fainter star in the stable pair among five dithering images.
An average value of deviation is 0.08 mag throughout the monitoring.
Therefore average of relative photometric error in this study is 0.1 mag.

Because reference stars are YSOs, we estimated the stability of absolute magnitude.
The \textsl{Ks} band magnitude of the fainter star in the stable pair, 2MASS J06125160+1758015, is 10.9 mag in the 2MASS catalog \citep{2MASS06}, while that calibrated from the magnitude of 2MASS J06125939+1757214 is 10.8 mag on average.
The difference of 0.1 mag should come from long-term variability of the two reference stars and we use this difference as an absolute magnitude error in this study.
We summarized our photometric results in Table \ref{tab:phot} in Appendix.

\section{Results}
\begin{figure}[p] 
  \begin{center}
    \includegraphics[width=8cm]{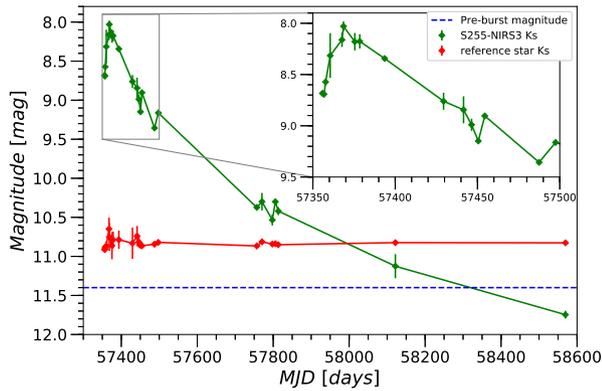} 
  \end{center}  
\caption{A light curve of S255-NIRS3 center object in the \textsl{Ks} band from November 29, 2015 to March 27, 2019. A zoomed-up plot around the peak flux is also shown at the upper right part in the original plot. In these light-curves, only relative magnitude errors are presented. There also exists systematic absolute magnitude of 0.1 mag. Sudden increase of flux is clearly seen towards December 12, 2015 (MJD=57368) and the following relatively shallow fading.}\label{fig:lc}
\end{figure}

Figures \ref{fig:field} (Top and middle parts) and \ref{fig:lc} show the observed image around S255-NIRS3 and the derived light curve of S255-NIRS3 in the \textsl{Ks} band.
In our observations, no new infrared sources were detected as compared with the UKIDSS archival images in 2009 \citep{Caratti17}.
Hence, S255-NIRS3 itself got brighter. Moreover, the surrounding nebulae also got brighter and this confirmed that the same origin made this whole flux increase, as shown from the previous NIR observations and light echo study in \citet{Caratti17}.

Our observations gave a detailed light curve of S255-NIRS3 in the NIR after the maser burst event with the highest cadence and the longest term. 
An average $1 \sigma$ error of magnitudes throughout the observations, including relative and absolute errors, are $\sim$ 0.2 mag as described in the previous section.
We derived a magnitude value of 8.7 mag in the \textsl{Ks} band on November 29, 2015 (UT), which is almost consistent with that of 8.9 mag on November 28 (UT) reported by \citet{Stecklum16}.
The slight difference ($\sim$ 0.2 mag) between the two epochs (28 and 29 November 2015) might be due to a larger uncertainty in the presented data or by the fact that the light-curve was steeply rising.
During the whole monitoring period, flux variation as large as 3.7 mag was detected (see Figure \ref{fig:lc}).

\section{Discussion}

\subsection{Characteristics of NIR variability of S255-NIRS3}
As seen in Figure \ref{fig:lc}, on the first date of observation, the magnitude of S255-NIRS3 was 8.7 mag, which is 2.7 mag brighter than that in its pre-burst phase obtained from the UKIDSS archival images in December 2009 (MJD $\sim$ 55173) \citep{Caratti17}.
The light curve showed a flux increase of approximately 0.7 mag until the beginning of December 2015 (MJD = 57368.591036), and then a relatively moderate fading of 3.7 mag until the last observation on March 27, 2019 (MJD = 58569.441348).
The magnitude at the last observation, 11.7 mag, became slightly below that at the quiescent phase, 11.4 mag.
This may be due to a larger uncertainty in the presented data or long-term variability of S255-NIRS3.
Additionally, week-by-week small-scale up-and-down flux variations were detected throughout the observations within the large-scale flux variation trend.
The drastic increase in flux further confirmed that S255-NIRS3 experienced an accretion burst event as previously indicated in \citet{Caratti17}.
In \citet{Caratti17,Caratti17MmSAI}, the authors claimed that the accretion event of S255-NIRS3 resembles EXor objects, low-mass YSOs possibly experiencing accretion burst events ($\Delta V \sim 2-3$ mag), but not so drastic as FUor objects ($\Delta V > 5$ mag) \citep{Audard14}, with its increase of bolometric luminosity. 
They also shown the obtained NIR spectra of S255-NIRS3 resembles EXor objects.
Another characteristic of EXor objects is the relatively short length of burst, typically a few years, while the length is generally longer than 10 years in FUor objects \citep{Audard14}.
In this study, the length of the burst is measured as about 2.5 years and this provides the last piece of evidence for confirming that the accretion burst in S255-NIRS3 is the first EXor-like accretion outburst observed in a MYSO.

While EXor objects frequently show the recurrence of a drastic flux increase \citep{Audard14},
 the monitoring observations at any wavelengths spanned only 4.5 years and only one event was detected in S255-NIRS3.
Further continuous of monitoring is essentially needed for searching the recurrence of the accretion burst in S255-NIRS3.

\subsection{Comparison of flux variation in S255-NIRS3 between maser and NIR}
\begin{figure}[p] 
  \begin{center}
    \includegraphics[width=8cm]{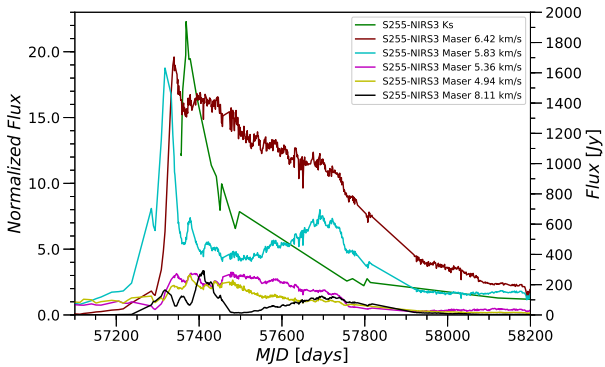} 
  \end{center}  
\caption{Light curves of S255-NIRS3 in the \textsl{Ks} band and 6.7 GHz methanol maser with major velocity components. The light curve in the \textsl{Ks} band is normalized so that flux on the quiescent phase (\textsl{Ks} = 11.4 mag) becomes unity. The normalized flux in the \textsl{Ks} band is shown at the left side of Y axis and the radio flux at the right side. The light curves of the bright maser components and infrared emission show a similar trend but their date of the peak fluxes differs by dozens of days.}\label{fig:lc_maser}
\end{figure}

\begin{figure}[p] 
  \begin{center}
    \includegraphics[width=6cm]{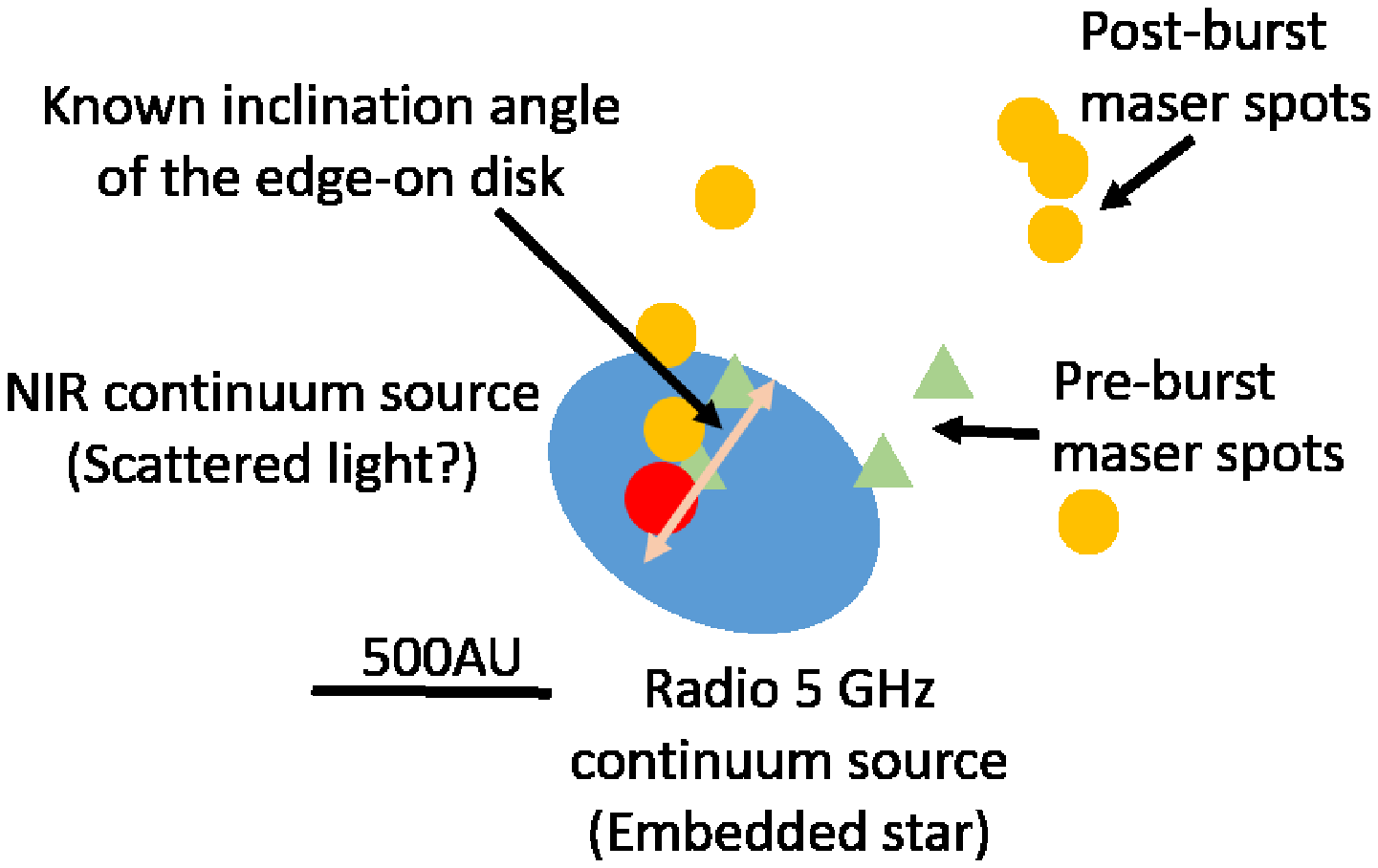}
  \end{center}  
\caption{Sky-plane image sketch of S255-NIRS3 region from radio observations including maser observations with VLBI \citep{Moscadelli17} and NIR observations \citep{Stecklum16}. NIR and radio emitting region are separated by approximately 1 light days in the plane of sky, and this is considerably shorter than the observed delay between the radio and NIR flux peaks.}\label{fig:position}
\end{figure}

Figure \ref{fig:lc_maser} compares the flux variation of major velocity components of the 6.7 GHz class II methanol maser (Sugiyama et al. in prep.) and the \textsl{Ks} band continuum emission.
Observations of the methanol maser were conducted at least daily during the whole NIR monitoring period.
Sudden increase in flux and its following moderate fading are also clearly seen in the light curves of the most and the next brightest maser velocity components (6.42 and 5.83 km/s).
The same trend in the maser variability was also reported in a previous radio monitoring work of \citet{Szymczak18}.
Table \ref{tab:peaks} summarizes the properties of flux variations in both the maser and NIR emissions.
Maser decay timescales are consistent with those seen in Figure 4 in \citet{Szymczak18}.
Note that maser emission is a stimulated emission and its magnification and timescale are not proportional to the change in physical conditions, such as radiation intensity or temperature of pumping.
As shown in Figure \ref{fig:lc_maser} and Table \ref{tab:peaks}, there are some similarities in the maser and NIR emissions: 1) the magnification of flux is large (larger than 10), and 2) the decrease in flux to the stable state is relatively slow (longer than two years).
These similarities indicate that the same origin, such as an accretion burst, causes the observed flux variations in both the NIR and the maser as previously indicated in \citet{Moscadelli17}.
The 6.7 GHz class II methanol maser is considered to be pumped up by mid-infrared emission from warm dusty disks at a temperature of about $\sim$ 120-300 K associated with MYSOs \citep{Cragg05,Sugiyama14}.
As per this hypothesis, when an accretion burst happens, maser luminosity should increase because the increase of stellar radiation warms up the dusty surrounding disk.
Note that the maser emission is stimulated emission and almost all the outbursting maser components have been still unsaturated \citep{Moscadelli17}, providing the condition that it does not simply linearly correlate with the stellar emission. Our observed trend also confirms this situation and therefore further supports the hypothesis that MIR emission from the disk pumped up the maser emission. 
The same suggestion was reported in a recent study with VLBI observations of the pre- and outburst phase of S255-NIRS3 \citep{Moscadelli17}.
The study demonstrated that spatial locations of maser spots at the post-burst phase spread more widely than those at the pre-burst phase.
Together with our NIR observation results, we made a sketch of the S255-NIRS3 referring their work and image as illustrated in Figure \ref{fig:position}.
This expansion suggests that increased radiation heats the disk and maser-pumping region expands toward the outside of the disk \citep{Moscadelli17}.

However, the peak flux observation date was 30-50 days earlier in the maser than that in the NIR, and also different among major velocity components of masers, as shown in the Table. \ref{tab:peaks}.
This time delay cannot be simply explained by the above mechanism, in which stellar radiation causes the increment of disk thermal emission.
However, assuming that the central NIR emission is scattered light as mentioned in \citet{Caratti17}, geometrical configuration of the scattered NIR emission may explain this delay.
The observed locations of S255-NIRS3 and maser emitting regions are displaced \citep{Stecklum16,Moscadelli17} by $<$ 0.1 arcsec, which corresponds to less than 180 AU in the plane of sky at the distance of 1.8 kpc.
This is approximately 1 light days, shorter than the observed delay of about 40 days, as shown in the Figure. \ref{fig:position}. 
Inclination of the disk might extend the actual distance; actually, previous MIR interferometric observations of a dust disk and VLBI observation of maser spots show a nearly edge-on inner disk inclination \citep{Boley13,Moscadelli17}.
Scattering at the flared disk edge or a stem of outflow cavity might explain the situation but observational evidence is not sufficient.

NIR observations started 100 days after the maser burst. Therefore, the observed flux peak in the NIR might possibly be the second peak after the accretion burst. If this is the case, the comparison of the NIR and maser peaks does not make sense.
The previous light echo analysis given by \citet{Caratti17} indicated that the accretion burst occurred from middle of June 2015.
If this date corresponds to the first flux peak in the NIR, time delay between NIR and maser could be about 3 or 4 months \citep{Fujisawa15, Caratti17}.
However, there does not seem to exist the maser peak corresponding to the second NIR flux peak, assuming that the maser peak appears 3-4 months behind of the observed NIR flux peak (MJD $\sim$ 57450-57500).
This hypothesis may be further examined with a multi-color light echo study of S255-NIRS3 in the future.
Immediate and high-cadence follow-up observations of burst MYSOs in both the NIR and maser right after burst events are critically important to reveal the true correlation between the NIR and maser, and a possible mechanism of the time-delay between them.

\begin{table}[p]
	\tbl{Property of flux variations in the radio maser emission and the NIR continuum emission}{%
		\begin{tabular}{cccc}
			& Maser (6.42 km/s) & Maser (5.83 km/s) & NIR (\textsl{Ks}) \\ \hline
			MJD at peak flux & 57339.5785 & 57317.6389 & 57368.5910  \\ 
			Flux at peak & 1705 Jy & 1632 Jy & 8.0 mag \\
			Flux at stable phase & 0.5 Jy & 60 Jy & 11.4 mag \\
			Decay timescale\footnotemark[1] & 408 days & 24 days & 61 days \\
			Magnification of flux\footnotemark[2] & 3400 & 27 & 23 \\
			\hline
	\end{tabular}}\label{tab:peaks}
	\begin{tabnote}
		1. timescale on that intensity decreases to half its peak.
		
		2. flux magnification from its quiescent phase before burst event.
	\end{tabnote}
\end{table}

\begin{ack}
	
This work has been supported by the Grant-in-Aid for Scientific Research (26$\cdot$10274) from the JSPS.

This publication makes use of data products from the Two Micron All Sky Survey, which is a joint project of the University of Massachusetts and the Infrared Processing and Analysis Center/California Institute of Technology, funded by the National Aeronautics and Space Administration and the National Science Foundation.

IRAF is distributed by the National Optical Astronomy Observatory, which is operated by the Association of Universities for Research in Astronomy (AURA) under a cooperative agreement with the National Science Foundation. 
\end{ack}

\bibliographystyle{apj}
\bibliography{apj-jour,ref.bib}

\clearpage

\appendix
\section*{Table of PSF fitting photometry results}

\begin{table}[ht]
	\tbl{Results of PSF fitting photometry in the \textsl{Ks} band}{%
		\begin{tabular}{ccc}
			MJD & S255-NIRS3 [mag] & Ref. star  [mag] \\ \hline
			57355.79 & 8.683 $\pm$ 0.023 & 10.908 $\pm$ 0.012 \\ 
			57356.77 & 8.692 $\pm$ 0.031 & 10.912 $\pm$ 0.024 \\ 
			57357.72 & 8.573 $\pm$ 0.033 & 10.883 $\pm$ 0.022 \\ 
			57360.51 & 8.315 $\pm$ 0.218 & 10.866 $\pm$ 0.061 \\ 
			57367.68 & 8.161 $\pm$ 0.070 & 10.649 $\pm$ 0.145 \\ 
			57368.59 & 8.030 $\pm$ 0.048 & 10.752 $\pm$ 0.178 \\ 
			57375.48 & 8.180 $\pm$ 0.085 & 10.862 $\pm$ 0.174 \\ 
			57378.48 & 8.176 $\pm$ 0.070 & 10.788 $\pm$ 0.105 \\ 
			57385.52 & 8.220 $\pm$ 0.030 & 10.757 $\pm$ 0.251 \\ 
			57393.57 & 8.343 $\pm$ 0.024 & 10.787 $\pm$ 0.117 \\ 
			57429.40 & 8.760 $\pm$ 0.083 & 10.832 $\pm$ 0.201 \\ 
			57441.55 & 8.844 $\pm$ 0.132 & 10.740 $\pm$ 0.130 \\ 
			57446.43 & 8.990 $\pm$ 0.059 & 10.817 $\pm$ 0.073 \\ 
			57450.52 & 9.148 $\pm$ 0.018 & 10.858 $\pm$ 0.024 \\ 
			57454.44 & 8.904 $\pm$ 0.007 & 10.864 $\pm$ 0.014 \\ 
			57487.45 & 9.356 $\pm$ 0.021 & 10.844 $\pm$ 0.015 \\ 
			57497.46 & 9.163 $\pm$ 0.004 & 10.820 $\pm$ 0.008 \\ 
			57738.62 & 10.401 $\pm$ 0.101 & 10.774 $\pm$ 0.231 \\ 
			57756.74 & 10.373 $\pm$ 0.020 & 10.867 $\pm$ 0.013 \\ 
			57770.45 & 10.299 $\pm$ 0.112 & 10.813 $\pm$ 0.035 \\ 
			57778.56 & 10.331 $\pm$ 0.076 & 10.679 $\pm$ 0.288 \\ 
			57797.65 & 10.534 $\pm$ 0.070 & 10.842 $\pm$ 0.014 \\ 
			57805.62 & 10.300 $\pm$ 0.014 & 10.837 $\pm$ 0.011 \\ 
			57813.49 & 10.420 $\pm$ 0.051 & 10.851 $\pm$ 0.015 \\ 
			58121.67 & 11.126 $\pm$ 0.154 & 10.824 $\pm$ 0.023 \\ 
			58569.44 & 11.747 $\pm$ 0.050 & 10.827 $\pm$ 0.026 \\
			\hline
	\end{tabular}}\label{tab:phot}
	\begin{tabnote}
		Only relative magnitude errors are presented. There also exists systematic absolute magnitude of 0.1 mag.
	\end{tabnote}
\end{table}

\end{document}